# WHAT $T_c$ CAN TEACH ABOUT SUPERCONDUCTIVITY

**Theodore H. Geballe**

Department of Applied Physics and Department of Materials Science, Stanford University, Stanford, California

**Gertjan Koster**

Geballe Laboratory for Advanced Materials, Stanford University, Stanford, California.

ABSTRACT. We compare the $T_c$s found in different families of optimally-doped High-$T_c$ cuprates and find, contrary to generally accepted lore, that pairing is not exclusively in the $CuO_2$ layers. Evidence for additional pairing interactions, that take place outside the $CuO_2$ layers, is found in two different classes of cuprates, namely the charge reservoir and the chain layer cuprates. The additional pairing in these layers suppresses fluctuations and hence enhances $T_c$. $T_c$s higher than 100K, are found in the cuprates containing charge reservoir layers with cations of Tl, Bi, or Hg that are known to be negative-$U$ ions. Comparisons with other cuprates that have the same sequence of optimally doped $CuO_2$ layers, but have lower $T_c$s, show that $T_c$ is increased by factors of two or more upon insertion of the charge reservoir layer(s). The Tl ion has been shown to be an electronic pairing center in the model system (Pb,Tl)Te and data in the literature that suggest it behaves similarly in the cuprates. A number of other puzzling results that are found in the Hg, Tl, and Bi cuprates can be understood in terms of negative-$U$ ion pairing centers in the charge reservoir layers. There is also evidence for additional pairing in the chain layer cuprates. Superconductivity that originates in the double "zigzag" Cu chains layers that has been recently demonstrated in NMR studies of Pr-247 leads to the suggestion of a linear, charge 1, diamagnetic quasiparticle formed from a charge-transfer exciton and a hole. Other properties of the chain layer cuprates that are difficult to explain using models in which the pairing is solely confined to the $CuO_2$ layers can be understood if supplementary pairing in the chain layers is included. Finally, we speculate that these same linear quasi-particles can exist in the 2-dimensional $CuO_2$ layers as well. It is possible that these particles will propagate chiefly in either the x or y direction and be appropriate candidates for fluctuating stripes and for d-wave superconductivity.

# INTRODUCTION

The transition into the superconducting state reflects the totality of all the underlying microscopic interactions in the system. As such, paying attention to the occurrence of $T_c$ throughout the Periodic Table of the Elements and its magnitude as a function of controlled external parameters such as pressure, composition, strain, dimension, and defects, is the most general approach for gaining an understanding of superconductivity. Discovering superconductors has been a fruitful enterprise for opening new fields of physics ever since Kamerlingh Onnes discovered back in 1911 that Hg loses all of its resistance abruptly just below 4.2 K [1]. For more than 4 decades superconductors were uncommon and poorly understood laboratory curiosities; there was no basis for predicting their occurrence and little connection with normal state properties. Hans Meissner found the barely metallic compound CuS to be superconducting whereas elemental Cu was not [2]. The rare and unpredictable occurrence of superconductivity, and the lack of an underlying microscopic theory, led Enrico Fermi at the University of Chicago in ~1950 to encourage two young colleagues, John Hulm and Bernd Matthias, to undertake a search for new superconductors. They soon found a number of new intermetallic alloys and compounds [3] and extended the work of Meissner, who in the 1930's had also found superconducting intermetallic borides [4]. In the same time period a parallel program was carried out in Russia by N.E. Alekseevskii and coworkers [5]. Whether there are still more bulk superconductors with novel properties remaining to be discovered is, of course, impossible to predict; but since there are now opportunities for synthesizing entirely new classes of materials and structures beyond phase equilibriums made possible by advances using thin film deposition and characterization techniques, there is good reason to believe that higher $T_c$s will be found.

   The Periodic Table was a valuable guide for predicting new superconductors particularly when Matthias noted that there is an amazingly simple dependence (known as Matthias' Rule [6]) of the magnitude of $T_c$ upon the average number of valence electrons per atom in elements and also in intermetallic compounds- i.e., $T_c$ is related simply to the electron density [7]. As a consequence of this "rule", superconductivity changed from being rare to being common. As the database increased, refinements were incorporated; superconductivity was found to be favored in specific structures [8], in particular in the A15 structure (also referred to as beta tungsten). This structure had the highest known $T_c$s up until 1986 as well as a number of other low temperature instabilities. It contains an unusual arrangement of non-intersecting chains of closely spaced transition metal atoms that impose features on the Fermi surface [9]. The discovery of the superconductivity of $V_3Si$ [10] and $Nb_3Sn$ [11] had a great impact, surpassed only by the discovery of cuprate superconductivity by Bednorz and Mueller more than 3 decades later, and led the way to new concepts in physics and a new high field-high current technology. The discovery of A15 superconductivity in $Nb_3Sn$ illustrates the fact that in searching for new superconductors, even though it may be a high risk endeavor, can have major consequences well beyond the original scope of the work [12]. In our opinion there is still much to be gained by continuing the research for new superconductors.

Little is known about the limits of superconductivity in the cuprates today. In this chapter we interpret the wide variation in $T_c$ that is found in different cuprate structures in terms of plausible intuitive models that are interesting in their own right, and might be of value in guiding paths to even higher $T_c s$.

## CUPRATE SUPERCONDUCTIVITY

There is nothing to compare with the discovery of High Temperature Superconductivity in the cuprates [13]. After two decades of intensive research there is no accepted theory nor is there consensus as to the superconducting pairing mechanism [14]. The cuprate charge carriers are highly correlated electronic systems that have sometimes been designated as "bad metals" [15] because Fermi-liquid theory is inadequate for treating the normal state properties.

Our aim in this chapter is modest. We start from the insulating side using a simple ionic model because we believe it provides a reliable way of gaining an intuitive understanding. The ionic model has long been used successfully for modeling insulating oxides and for understanding their magnetic properties. It is a limiting case of very strong correlation and thus has credibility as an initial approximation for the nearly insulating cuprate superconductors. In the Born approximation the large attractive Madelung energy is balanced by the repulsive overlap energy. We have found it useful, following Moyzhes and Suprun *et al.* [16] to modify the Born's equation by using the high frequency dielectric constant to account for the electronic polarizability of near neighbor ions and the low frequency dielectric constant to account for the ionic polarizability of more distant ions. This modification leads to a statistically significant classification of a large number of oxides, and predictions as to their stability, metallicity and instability to within about 1 eV [17]. The modified Born equation is a simplified representation of the local density (LDA) approximation [18].

### *Pairing and $T_c$s in the cuprates*

All the cuprates with high $T_c$s contain 2-dimensional layers of $CuO_2$ that upon sufficient doping become superconducting fluctuations. $T_c$ is found to increase when the number of $CuO_2$ layers per unit cell increases from $n = 1$ to 3. This is not surprising because the close spacing of the $n$ layers within the unit cell can be expected to stabilize the 3 dimensional fluctuations by quantum tunneling [19]. The decrease in $T_c$ with further increases in $n$ is discussed later. However focusing exclusively on the $CuO_2$ layers cannot explain some significant variations in $T_c$ that are found in structures that have the same sequences of $CuO_2$ layers but have different intervening layers, and that is the subject we address here.

The Cu ion;

Before proceeding to discuss the cuprates we recall some facts that make the Cu ion unique. In the vapor phase $Cu^{+2}$ has the highest 3[rd] ionization potential of the transition metals. This large energy is retained in the condensed state as is evident from the electrode potentials of ions in aqueous solution [20]. Electrode potentials provide rough estimates of the relative ionic energies in crystalline oxides because in both the aqueous and crystalline environments the cations are coordinated by oxygen ions. The standard electrode potential for charge transfer $Cu^{+3} + e^- = Cu^{+2}$, $E(0) = +2.4$eV is very high. It follows that in cuprates the doped holes will reside mainly on oxygen sites (in contrast to other transition metal oxides where the cations are oxidized upon hole doping). On the other hand, the standard electrode potential for the reaction $Cu^{+2} + e^- = Cu^{+1}$ is quite low, $E(0) = -0.15eV$, showing that $Cu^{+2}$ can easily coexist with $Cu^{+1}$. Consequently in the condensed state $Cu^{+1}$ and $Cu^{+2}$ are close in energy which translates in the Hubbard model to a moderate $U$ that splits a half-filled narrow band due to the on-site coulomb (Hubbard $U$) repulsion. Of course in the crystalline cuprates, the crystal field states, the band, exchange and correlation energies must be included. But the ionic energies are the largest so we can assume without further calculations `that $Cu^{+3}$ ($d^8$ configuration) does not play a significant role in the dynamics of the cuprates with the consequence that the cuprates are "charge transfer insulators" rather than Mott insulators [21].

In the undoped parent compounds the $CuO_2$ layers are insulating antiferromagnets containing $Cu^{+2}$ ions. The Cu is in a $d^9$ state; 2 of the 4 electrons needed for charge neutrality in the $CuO_2$ layer come from other layers such as the La layers in La-214. The overlap of the half filled Cu ($x^2$-$y^2$) d-levels results in a narrow band that, when $U$ is > the band width, splits into the upper and lower (Hubbard) bands [22]. In this chapter we restrict the discussion to the doping of holes in the $CuO_2$ layers that can be achieved by substitution of a cation with a lower valence (e.g., Sr for La), or by cation valence reduction (e.g., $Tl^{+3}$ to $Tl^{+1}$), or by the addition of negative oxygen ions. Upon hole-doping, the charge resides mainly on the oxygen site and the antiferromagnetism is rapidly destroyed. Concentrations >0.05 holes per Cu become superconducting [23]. A dome shaped curve [24] is found for all the cuprates. It is commonly assumed that the dome shape is universal with the maximum $T_c$ found at an optimum doping concentration $p$=0.16 holes/Cu. However, it is obvious that the 3-dimensional superconducting condensation measured by $T_c$ depends upon coupling between the layers and, because as we argue below the coupling is not universal, and therefore we should not expect there to be a single concentration for which $T_c$ becomes optimum. Thus the results of Karppinen that $T_c$ of Bi-2212 occurs for $p$=0.12 (see below) should not be surprising [25]. However we believe that the comparison of the optimum $T_c s$ in the different families of cuprates, Table 1, is meaningful. $T_c$ increases across the rows for $n$ = 1, 2, 3 (and decreases for $n > 3$) for reasons that will be discussed later. What concerns us here are the differences in the $T_c s$ of optimally doped cuprates that have the same number, n, of $CuO_2$ layers in the unit cell (the columns in Table 1). In order to account for these strong variations in $T_c$ it is necessary to assume either that the superconductivity of those with the higher $T_c s$ is enhanced, or that the superconductivity in the $CuO_2$ layers of the cuprates with the lower $T_c s$ is depressed, or, that both effects are present.

Table 1: Variation in $T_c$

| $CuO_2/_c$ | $n$=1 | | $n$=2 | | $n3$ | |
|---|---|---|---|---|---|---|
| | $T_c$(K) | Separations (Å) | $T_c$(K) | Separations (Å) | $T_c$(K) | Separations (Å) |
| LSCO-214 | 40 | 6.6 | - | - | - | - |
| Hg-12($n$-1)$n$ | 98 | 9.5 | 127 | 9.5 | 134 | 9.5 |
| Tl-12($n$-1)$n$ | - | | 103 | - | 133 | - |
| Tl-22($n$-1)$n$ | 95 | 11.5 | 118 | 11.5 | 125 | 11.5 |
| Bi-22($n$-1)$n$ | 38 | - | 96 | - | 120 | - |
| Y123 (6 GPa) | - | - | 95 | 7.9 | - | - |
| Y124 (6 GPa) | - | - | 105 | 9.8 | - | - |

Many mechanisms are known to depress $T_c$, including the competition with other kinds of long range order, local site disorder, and structural deformation (e.g. layer buckling). Competition with commensurate or nearly commensurate density waves causes large decreases in $T_c$. For example large dip in $T_c$ for $x$ = 1/8 in $(La_{1-x},Sr_x)_2CuO_4$, or its complete destruction in $(Nd_{1-x},Sr_x)_2CuO_4$ [26], and $(La_{1-x},Ba_x)_2CuO_4$ [27] at the same 1/8 doping level. However there has been no evidence for competitive ordering in optimally doped $(La_{1-x},Sr_x)_2CuO_4$ that can account for the ~50K reduction in $T_c$. Disorder is also known to depress $T_c$ but again in smaller magnitudes than needed. The $T_c s$ of the Bi-cuprates are somewhat lower than in the corresponding Hg and Tl cuprates (Table I). It is possible that the Bi cuprates are more disordered or that the negative-$U$ Bi ions are somewhat less effective pairing centers. Typically there are excess $Bi^{+3}$ that replace $Sr^{+2}$ next to the apical oxygen. When that disorder is replaced by a less intrusive disorder by substituting $Y^{+3}$ for $Ca^{+2}$ between the $CuO_2$ layers, Eisaki *et al.* [28] find $T_c$ increases from 90 to 96K.

## INTERACTIONS BEYOND THE CUO$_2$ LAYERS:

We first present plausible evidence that interactions outside the $CuO_2$ layers must be taken into account. We examine $T_c s$ in two classes of cuprates: i) The first consists of the charge reservoir layer cuprates from which we infer that there is an electronic pairing mechanism involving the negative-$U$ center ions; ii) The second has layers consisting of quasi 1-dimensional double-chains of CuO (sometimes described as "zigzag" chains) in which pairing is found. In the next sections we consider structural, compositional, pressure-dependent transport, and NMR/NQR data in the literature that collectively provide persuasive evidence for the enhancement hypothesis in both the charge reservoir and chain layer cuprates.

### *Pairing centers in the charge reservoir layer cuprates*

The charge reservoir layer cuprates contain additional layers of the oxides of the ions of Tl, Bi that are well known in aqueous solution and in solids as well [20] to have unstable paramagnetic $6s^1$ configurations that disproportionate to form diamagnetic ions with $6s^0$ and $6s^2$ configurations. Hg exists in solution as a two-center diamagnetic ion that exchanges charges in units of two just as the Tl and Bi ions do.

In solids the ionic picture must of course be modified but, as already noted, the ionic energies found in aqueous solution are large and useful initial approximations for insulating oxides. The three heavy Hg, Tl and Bi ions all have the paramagnetic $6s^1$ configurations that are unstable. In the cuprates that have the highest $T_c s$ (> 100K) these ions are in the so-called charge reservoir layers, separated from the $CuO_2$ layers by the apical oxygen ions in the alkaline earth-oxide layers. The charge reservoir layers are so named because of their ability to dope the $CuO_2$ layers. We propose that they have the additional important function of providing negative-$U$ pairing centers and it is more accurate to call them negative-$U$ charge reservoir layers. The $T_c$ of Hg-1223 under pressure reaches the highest recorded $T_c \sim 160K$ [29]. Comparison with other cuprates that do not contain negative-$U$ center ions provides convincing evidence that the negative-$U$ ion layers, directly or indirectly, are responsible for enhancement of $T_c$. Our interpretation is that the enhancement is due to interactions with the charge reservoir layers as we now show.

In the well known 214 family of cuprates (based upon $La_2CuO_4$) $T_c$ reaches a $T_{c,max} \sim 40K$ when the $CuO_2$ layer is optimally doped with Sr; $T_c s$ above 50K can be reached in strained epitaxial thin films when the doping occurs with the insertion of oxygen interstitials to form the staged compound $La_2CuO_{4+x}$ [30] and up to 45 K in oxygen doped $_c$ samples [28], but there is no evidence for higher $T_c s$ in the 214 family.

What must be addressed is the mechanism by which the $T_c$ of the "optimally" doped 214 compound is increased to >90K by inserting charge reservoir layers containing ions of Tl, Hg or Bi and oxygen. As can be seen in the Table 1 the 6.6Å distance between the $CuO_2$ layers in the 214 compound is increased by another 5Å by the insertion of TlO layers between them , a change that by itself would be expected to decrease $T_c$.

In our model the pairing induced by the negative-$U$ ions enhances $T_c$. We have considered the possibilities that either a given negative-$U$ ion acts as a resonant pair tunneling center, or that clusters of negative-$U$ ions become coherent with CuO$_2$ layer, or that the negative-$U$ ion layer itself develops an independent 2D order that subsequently becomes coherent with the CuO$_2$. We are not aware of any experiment that might distinguish them, however theoretical considerations [31] rule out the likelihood that of there being independent two dimensional order parameters. It is our opinion that the most likely case is the intermediate one where clusters of fluctuating negative-$U$ ions form and then utilize the most favorable sites for condensation. This latter possibility gains support in the theoretical model of a structure consisting of negative-$U$ centers in the barrier of Josephson-junction weakly connected to the electrodes [32]. An interesting prediction of that model is that there will be a strong enhancement of I$_c$R the product of the critical current and the normal resistance in the c direction.

*Negative-U center electronic pairing in a model system*

Anderson [33] introduced the concept of a negative-$U$ center to explain the failure to observe EPR signals in chalcogenide glasses and simultaneously the pinning of the Fermi energy. He noted that the lattice relaxation around a localized electron could overcome the repulsive coulomb (Hubbard $U$) energy of adding a second electron, and thus result in an effective negative-$U$. A further analysis by Moyzhes and Suprun [16] showed that in PbTe doped with valence skipping ions the electronic response (i.e. the polarization) of the surrounding medium can over compensate the coulomb repulsion. The relaxation of the polarization charge, set by the high frequency dielectric constant in PbTe, is large, ~30. In the model (Pb,Tl)Te system experiments support an electronic superconducting pairing mechanism as we now briefly discuss.

PbTe in which ~ 1% of the +2 Pb is replaced by Tl has long been a good model system for studying negative-$U$ center induced superconductivity [34]. Recent investigations by Matsushita, *et al.* [35], and by Schmalian *et al.* [36] have illuminated the role of Tl ions. For low concentrations < 0.3% Tl acts like a shallow acceptor forming $Tl^{+1}$ as evidenced by the Hall constant that shows one hole is doped into the valence band per added Tl ion. Above ~0.5%, however, the Hall constant becomes more nearly independent of the doping; the Fermi level is pinned as a result of disproportionation reaction: $2Tl^{+2}$ $(6s^1) \rightarrow Tl^+1$ $(6s^2)$ + $Tl^+3$ $(6s^0)$.

The near degeneracy of the +1 and +3 states is suggested by a systematic study of the temperature and field dependence of the resisitivity. The data can be fit by the charge-Kondo model that requires that the two charge states of Tl ions be degenerate within ~ $k_B T_c$ [35]. Kondo-like behavior and $T_c$ set in at nearly the same concentration where the Hall effect indicates the Fermi level is pinned. The superconducting transition is driven by the gain in energy when the pairing on the different Tl ions becomes coherent presumably by interacting through the valence band states.

The lesson taken from the above (Pb,Tl)Te investigations is that for pairing to occur in the charge reservoir layers of the cuprates, the negative-$U$ ion configurations of at least some of the centers must be nearly degenerate (within $k_B T$) in energy. If the doping from the charge reservoir layers is achieved by a change of valence of the negative-$U$ ion, the Fermi level is automatically pinned and the condition for degeneracy is assured. A possible approach for discovering new superconductors is to identify structures in which the negative-$U$ ions can be incorporated. Potential negative-$U$ ions are listed by Koster *et al.* [17]. The difficult step is to find those systems where the chemical potential can be adjusted so as to bring the two levels into degeneracy. In terms of the Emery-Kivelson model [37] optimal doping is determined by the intersection of the pair amplitude and the stiffness as a function of superfluid density, see Fig. 1. For the 214 experiments show the optimal density occurs at 0.16 and $T_c$=40K. Assuming the optimum is at 0.16, as is frequently done, the doubling of $T_{c,optimum}$ would require both the stiffness and the pair amplitude to change accordingly. However, the introduction of pairing centers in layers between the $CuO_2$ layers will mostly suppress fluctuations and consequently will shift the occurrence of $T_{c,optimum}$ to a lower superfluid density as shown schematically in Fig 1. In fact experimental evidence shows that $T_{c,optimum}$ for the Bi-2212 does occur near the 1/8 concentration as will be discussed below [25].

<u>Thallium cuprates</u>

T. Suzuki *et al.* have observed by XPS spectroscopy that the Tl 4f7/2 core level valence in Tl-2223 lies roughly midway between the core levels of the $Tl^{+3}$ ($Tl_2O_3$) and $Tl^{+1}$ ($Tl_2O$) reference standard oxides. This indicates the presence of degenerate $Tl^{+1}$ and $Tl^{+3}$ states that can exchange pairs of electrons with the $CuO_2$ layers [38].

The $T_c s$ of ceramic samples of the (Cu,Tl)-1223 and Tl-1223 cuprates as prepared are ~ 100K and increase monotonically up to 133K upon annealing temperatures up to 550 °C in vacuum [39]. The 4f 7/2 core level of the Tl ion in the TlO charge reservoir layers in the as prepared samples are rather broad and are centered around the peak of $Tl^{+3}$ in the reference compound. Terada *et al.* find the peak shifts to midway between the values of the $Tl^{+1}$ and $Tl^{+3}$ references upon the annealing. It is our interpretation that the shift is due an increased presence of $Tl^{+1}$ (as we have argued above the $Tl^{+2}$ paramagnetic configuration is at higher energy); however, the spectra have not been fully analyzed [40].

<u>Mercury Cuprates</u>

The mercury cuprates are interesting for several reasons beyond having the highest known $T_c >$ 160K found in the Hg-1223 compound under pressure [29]. The homologous series $HgBa_2Ca_{n-1}Cu_nO_{2n+2+\delta}$ has been synthesized [41] all the way from $n$ =1 to $n$ =7 with $T_c$ for the optimally doped samples rising from 97K for $n = 1$ to a max at $n = 3$ and then falling at a decreasing rate with further increase in $n$ until $T_c = 80$ K for $n = 7$.

The increase from $n = 1$ to $n = 3$ is common to all the cuprates (Table 1) and follows from the coupling of the layers by quantum tunneling [42]. The maximum observed for n=3 and subsequent decrease with higher $n$ can be understood from NMR investigations that show that the $CuO_2$ layers are not uniformly doped [43]. In optimally doped n=5 samples the inner layers are antiferromagnetic, $T_N = 60$K with ~0.35 $\mu_B$/Cu [44] Cu while the outer layers are superconducting $T_c$=108K. The robust persistence of the superconductivity as evidenced by a negative curvature of the dependence of $T_c$ upon $n$ for $n > 3$ would be difficult to understand if the superconducting interactions were confined to the single outer $CuO_2$ layers, see Fig. 2.

Mukada *et al*. [45] find for an $n = 5$ underdoped sample, that the three inner layers are anti ferromagnetic, ($T_N$= 290K with 0.68 $\mu_B$/Cu) while the two outer layers are both antiferromagnetic (0.1 $\mu_B$/Cu) and superconducting, $T_c$= 72K. In a somewhat comparable structure, but one without negative-$U$ centers, Bozovic *et al*. [46] find that an isolated 1 unit cell thick film (two $CuO_2$ layers) of optimally doped $(La_{1-x},Sr_x)_2CuO_4$ (sandwiched between antiferromagnetic films of undoped $La_2CuO_4$ by sharp interfaces have $T_c$s of only 30K. The much higher $T_c$s ~ 80K found for the $n$ >5 Hg cuprates that have single layers of doped $CuO_2$ interfaced on one side with AFM layers and on the other with BaO-HgO layers make it plausible that the latter layers are contributing to the pairing

The possibility has been raised that the $T_c$s in the Hg cuprates are unusually high because the layers are flat (O-Cu-O bonds are 180 degrees), which is known to be detrimental to $T_c$ [47,48]. However, these ideas are at odds with the neutron diffraction data for Hg-1212 where under pressures of ~100kbar the $CuO_2$ layers become buckled to the same extent found in the 214 cuprate and the apical oxygen moves closer to the planes, while the $T_c$ increases [Fig 3]. As noted by Jorgensen [49] if the buckling could be prevented $T_c$ should be even higher. While to date the only the Hg-1212 diffraction patterns has been studied under pressure it seems likely that the buckling under pressure will occur in all the Hg cuprates because the pressure dependence of $T_c$ scales for the $n = 1, 2$, and 3 as shown in Fig 3.

The reason $dT_c/dP$ is constant = 2.0 K/GPa from low doping levels to optimum doping [50], as shown in Fig. 3, is not obvious, and certainly does not follow from models that assume the changes are due to charge transfer. Such models would reasonably expect to find a steadily decreasing pressure coefficient from low to optimal doping. However, the behavior may be consistent with a negative $U$ model because pressure should increase the overlap of the pairing centers in the HgO layers with the $CuO_2$ layers. Raman data suggests that the overlap is through the apical oxygen ions. [51,52].

The HgO-BaO layers are highly disordered. There are a large number of oxygen vacancies in the HgO layers. XAFS measurements of the Hg-Hg distances are of such poor quality that they cannot be modeled [53]. Consequently the negative $U$-ion is probably a more complex entity than the idealized two center ion. In the model (Pb,Tl)Te system it is estimated that only a few percent of the Tl negative $U$ ions are pairing centers [35], thus it is not unreasonable that substitution of a substantial concentration of Re or Cu cations substituted for Hg layers has little effect on $T_c$. Experiments to determine the Hg valency such as has been done for the Tl cuprates (see above) would be helpful. For this purpose better and larger Hg cuprate single crystals are becoming available [54].

The Bismuth Cuprates

The bismuth cuprates have $T_c$s that are somewhat lower than the corresponding Hg and Tl cuprates (Table 1). This suggests that either the Bi negative-$U$ centers are not such effective pairing centers, or that their enhancement is counteracted by a competing effect. Disorder is not an unreasonable possibility because there is known to be considerable anti-site disorder and excess Bi on the Sr sites [55]. Perhaps even more important is the incommensurate superstructure found in the BiO layers [28,56,57] that causes displacements throughout the unit cell including large amplitude waves of CuO buckling along the a-axis [55]. The inhomogeneous images observed by STM [58,59] are also evidence of disorder although it is not obvious how much of the observed disorder may be due to the surface layer reconstruction because the tunneling must be through the orbitals extending from the surface.

The investigation of Karppinen *et al.* [25] that utilizes independent electrochemical and spectroscopic means of analysis and finds them to be in agreement has two significant findings. First, half of the charge introduced by substituting Sr for Y on sites between the two $CuO_2$ layers in Bi-2212 ends up in the non adjacent BiO layers. The second is that the $T_{c,\text{opt}}$ for Bi-2212 occurs when the carrier concentration in the $CuO_2$ layers is 0.12 (see Fig. 5). This is most significant because it is near the same 1/8 concentration where, it is well known from experiments on cuprates that do not contain charge reservoir layers, that $T_c$ is depressed or nonexistent because charge ordering and static stripe formation compete successfully with superconductivity [60,61].

As mentioned above, the Emery-Kivelson model [37] predicts that the intersection of the superfluid density curve when the phase fluctuations are stiffened, and $T_c$ optimum will occur at lower doping levels (as sketched in Fig 1).

*The chain layer cuprates*

Many investigations have been carried out in the single and double chain cuprates that lead to the conclusion that the chain layers support pairing. We now consider three different chain-layer cuprate structures that are comparable insofar as their layering sequence is concerned, but differ in the structure of the chain layers themselves. The chain layers consist of either single CuO chains, or double ("zigzag") CuO chains, or a combination of alternating single and double layers as shown in Fig. 6 [62]. In all three structures the quasi-one dimensional chains are separated from the blocks of $n$=2 $CuO_2$-(Y,Pr)-$CuO_2$ layers by layers of BaO. Evidence given below suggests that the chains that run in the b direction interact with each other in the a-direction indirectly via the $CuO_2$ layers. An important difference is that the oxygen ion concentration in the single 123 chain layers is variable whereas in the 124 (both the 124 and 248 notations are used interchangeably in the literature and we do likewise) it is fixed. This permits doping over a wide range in the $CuO_2$ layers of the 123 cuprates, but not in the 124 cuprates. In the non-stoichiometric 123 cuprates the vacancy diffusion leads to various kinds of short and long range ordered structures [63], whereas the 124 cuprates are stoichiometric and the diffusion is very much slower. Doping on the $CuO_2$ layers of course is possible by cation substitution on the Y site.

Evidence from nuclear quadrupole resonance in the double chains

The NQR investigation of Sasaki *et al*. [63] provides direct evidence that the superconductivity discovered by M. Matsukawa *et al*. [64] in Pr-247, originates in the double chains layers. As can be seen in Fig. 6, the 247 structure is composed of alternating Pr-123 and Pr-124 units. Neither of the units by themselves has been found to be superconducting and, as initially prepared by sintering Pr-247, also is not superconducting. However it becomes superconducting with zero resistance at $T \sim$ 10K when annealed in vacuum at 400 °C. The NQR Cu resonances associated with the four different Cu sites in the Pr-247 structure are well resolved allowing them to be followed separately. The $CuO_2$ layers order antiferromagnetically around 280K (See Fig.7b) as they do in Pr-123 and Pr-124. The relaxation data observed in the Pr-247 samples provide the evidence that the superconductivity resides in the double chain layers. As can be seen in Fig. 7a, near $T_c$ the temperature dependence of the nuclear relaxation rate of the double chain Cu nuclei changes markedly from Tomanaga-Luttinger one dimensional behavior to a reduction in density of the electronic states. This can hardly be a coincidental, and must be due to the superconductivity.

While ~10K may not be "high temperature" in comparison with other cuprates it is very high when compared with other comparable 1D systems such as the polymer $(SN)_x$ [65]. The fact that it originates in a cuprate where the $CuO_2$ layers are insulating and antiferromagnetic [78] is significant; and leads us to suggest the existence of the linear diamagnetic quasi particles discussed below.

At this time we offer no explanation for how the annealing turns on the superconductivity. The temperature dependence of the resistivity before and after annealing gives evidence for transport by parallel conduction paths. The annealing increases the room temperature resistance presumably due to the single chains becoming insulating. Upon cooling there is a striking increase in the conductivity of the annealed sample that culminates in the superconducting transition. Comparable annealing experiments of the Pr-124 double chain cuprates show no such effects. In order to account for the 1D transport and superconductivity we suggest the formations of a linear diamagnetic bound exciton hole (eh) quasi particle (Fig. 8c) that is discussed below.

Evidence from anisotropy

The CuO chains running in the b-crystal direction are directly or indirectly responsible for the considerable planar anisotropies observed in *d.c.* and optical conductivities in the normal states of Y-123 and Y-124, and in their penetration depths in the superconducting state. Basov *et al.* [66] find from far infrared data that the planar anisotropies, in agreement with transport data, are large and temperature independent. At room temperature $\sigma_b/\sigma_a = 1.8$ [67] in the Y-123 and in the Y-124 it is even larger ~3 [68]. Corresponding penetration depth measurements find rather interestingly that the anisotropy of the superfluid densities are almost the same. If it is assumed that the anisotropy is simply due to the orthorhombicity of the $CuO_2$ layers, then why is it greater in the 124 when the 124 is less orthorhombic? If it is attributed to a proximity induced superfluid density on layers and metallic CuO chains [69,70] then it is fails to predict the observed wide range over which the anisotropy is temperature independent [66]. However the experiments are consistent with models that assume intrinsic pairing in the chain layers.

Evidence from pressure

Superconducting pairing in the double chains is a likely explanation for why the $T_c s$ of Y-124 rise above those in the single chain Y-123 as the pressure is raised. The stoichiometric double chain Y-124$O_{15}$ cuprates are underdoped, $T_c$ = 80K at atmospheric pressure, whereas $T_c$ for the non-stoichiometric optimally doped Y-123$O_{6.93}$ 93K. However $T_c$ for Y-124$O_{15}$ rises to 108K at 6 Gpa exceeding the $T_c$ of optimally doped Y-123 at any pressure, a result that does not follow from any proximity effect theory. The abnormally large increase in $T_c$ of the Y-124 with pressure might in some part be due to additional charge transfer; however charge transfer does not explain the abnormally large anisotropic strain dependence, of $T_c$ of the Y-124. Strain dependence in the a-direction (when the chain-chain planar distance is reduced) [71,72] is much larger than in the c direction (when the layer spacing is reduced that should affect charge transfer) suggesting that chain-chain coupling plays a key role. Evidence from Zn doping discussed below indicates that the coupling is made through the CuO$_2$ layers.

Evidence from cation substitution in the n=2 of CuO$_2$ layers.

If the reasonable assumption is made that substituting Pr for Y between the CuO$_2$ layers has the same effect in 123 and 124, then there is further evidence for pairing on the double chains. For instance, (Y$_{0.4}$Pr$_{0.6}$)123 is not superconducting while the same composition in the 124 has $T_c \sim$ 50K, a result that is difficult to explain by any proximity effect [73].

Zn is known to dope in the CuO$_2$ planes and destroys $T_c$ rapidly in all the cuprates [74]. In the Y123 and Y124 cuprates $T_c$ decreases rapidly and roughly at the same rate with Zn substitution. This alone might suggest that all the superconductivity is in the $CuO_2$ layers, but such an interpretation is not viable in the light of other evidence such as the large d$T_c$/d$a$ mentioned above [75]. Evidence we now cite suggests that the double chains interact with each other by coupling with each other indirectly through the $CuO_2$ planes. In non-superconducting Pr-124 the b axis (chain direction) shows good metallic conductivity, the a-axis resistivity peaks around 140 K. The results can be modeled by parallel paths of the conducting chains and the semiconducting $CuO_2$ planes; the planar transport anisotropy is 1000 at 4 K [76]. Upon doping with Zn the material becomes insulating along the a-axis while the b-axis continues to show metallic behavior [77]. The disappearance of the Fermi level as found in an ARPES investigation [78] is explained by the increased one dimensionality of the double chains They presumably become more decoupled when the coherence length in the $CuO_2$ layer is destroyed by the Zn and a competitive 1d instability such as a charge density wave becomes the ground state. The coupling between the chain layers and the $CuO_2$ layers is likely to be through the apical oxygen for which there is independent evidence [12]. More complete literature references are given in the chapter by Valo and Leskela [79].

*Other chain layer compounds*

<u>The ladder compound</u>

A comparable double "zigzag" CuO chain arrangement to that found in the 124 cuprates is found in $Sr_{14-x}Ca_{x-12}Cu_{24}O_{41}$ (14-12-24-41) which undergoes a broad superconducting ($T_c$~10K) under pressure [80]. The structure contains alternating layers of single chain $Sr_2CuO_3$ and double chain $SrCuO_2$ layers. The double chains are separated in each $SrCuO_2$ layer by the rungs of two leg ladders. There is evidence that the superconductivity originates in the $SrCuO_2$ layers that is due to charge transfer from the single to the double chain layers [81].

A theoretical model predicted the superconductivity in the 14-12-24-41 structure prior to discovery [82] by assuming that the pairing occurs in the 2-leg ladders and confirming experimental evidence has been found [83]. In this model the spins of the double chain coppers are assumed to be connected by ferromagnetic superexchange via the oxygen $p_x$ and $p_y$ orbitals. The fact, that the same double chain configuration in the Pr-247 cuprate becomes superconducting rather than ferromagnetic, shows that subtle differences in coupling or doping can result in major changes in the ordering of quasi one-dimensional systems.

<u>Finite chain lengths.</u>

Infrared studies show that there is no anisotropy in the single chain cuprates for oxygen concentrations < 6.65 per unit cell, or for chain lengths < 15 to 20 Å for which $T_c \sim 60K$ [84]. The authors show that for higher doping when the chain length fragments exceed 20 Å there is a marked change in properties. The electromagnetic response in the normal state becomes coherent and quasi one-dimensional. Correspondingly the superfluid density in the b direction grows rapidly while in the a-direction it remains flat. As pointed out above, proximity effect models have difficulty in accounting for the identical temperature dependences over a wide temperature range in the a- and b-direction [66]. Strain- dependent measurements in the single chain cuprates are ambiguous because of the oxygen mobility allows for different oxygen ordering on the chains [85].

### *Superconductivity originating in the CuO$_2$ layers*

We speculate that the linear spinless charge one quasiparticles that explain the superconductivity of the Pr-247 quasi 1-dimensional double chains may equally well exist in the CuO$_2$ layers of all the cuprates. In the limit of negligible oxygen–oxygen near neighbor hopping ($t_{pp}$) such a quasiparticle model leads naturally to fluctuating stripes and d-wave superconductivity. These considerations lead to the prediction that if a 2-dimensional CuO layer (i.e., a structure where the vacant sites in the CuO₂ layer are filled with Cu) could be synthesized and properly doped it could have double the number of quasi particles than presently known cuprates [86].

<u>The exciton-hole (eh) quasi particle</u>

Low-energy charge-transfer excitations involve the transfer of electrons from the highest-lying oxygen level to the upper Hubbard band [21, 86]. They are estimated to be $\leq 2$ eV for La₂CuO₄ in the antiferromagnetic state at low temperatures [87], and in the same energy range HgBa₂CuO₄ [54] and likely all the high $T_c$ cuprates. There is also considerable subgap structure. The lowest peak at ~ 0.4 eV is in reasonable agreement an ionic estimate of the lowest energy charge transfer exciton [86] taken to be the gap energy less the screened interaction between the bound charges giving an energy $E_{ex} = E_g - q^2/\varepsilon r$. Here, $q$ is the absolute value of the charges, $r$ is their separation, and $\varepsilon$ is the dielectric response. Putting $E_g = 2$eV and $r = 2$ Å and making the reasonable assumption that for the short distance, $\varepsilon = \varepsilon_\infty = 5$, gives $E_{ex}$ +0.5 eV. Some of the subgap structure may also be due to multi-magnon/phonon processes [88].

Upon doping the bands broaden and the gap edge is lowered [54]. In fact RIXS data [89] suggest that that the spectral weight of the lowest lying exciton in the undoped compound is transferred to the continuum intensity below the gap, In our model this occurs when the charge transfer exciton combines with the doped hole to form a bound exciton hole (eh) quasi particle. The various ionic configurations to be considered in the CuO₂ layers are shown in Fig. 8.

For reasons given earlier doped holes mainly reside on the oxygen sites (see Fig. 8b). In the ionic model a low energy singlet is possible when the hole is attracted to the polarization cloud of the lowest lying charge transfer exciton (Fig. 8a) resulting in a new quasi particle that we call an eh (exciton-hole) particle (Fig 8c). The eh particle is a linear charge-one spin-zero quasi particle with an electrostatic energy $E = E_{ex} - [q^2/\varepsilon r - q^2/2\varepsilon r] = +0.5 - 0.72$eV$= -0.2$eV. The well-known Zhang Rice (ZR) singlet is an alternate configuration that places the doped hole in a symmetrical molecular orbital the oxygen ions surrounding a given Cu ion [90] and is stabilized by exchange energy [91]. However the eh-singlet is stabilized by Coulombic energy [78] and more importantly is a favorable configuration for stripe formation. A bent configuration rather than linear configuration would also have higher energy due to interaction $V_{pp}$, between the oxygen ions, (Fig 8e) [92].

In the limit $t_{pp} = 0$ the electron dynamics are purely one dimensional as depicted in Fig 8 c) [92]. Cluster calculations, however, suggest that $t_{pp}/t_{pd}$ is in the range of 0.3 [93] raising some question about the validity of one dimensional transport. The eh-particle, (Fig 9 d) will be dressed; in fact the extended version of the eh-particle (fig 8e), in which $t_{pp}/t_{pd}$ should be close to zero, has an even lower coulombic energy [94]. At the higher temperatures however entropy will favor the eh-particle.

The ionic version of the phase diagram in Fig. 1, is qualitatively consistent with generally accepted phase diagrams [95] except that we have allowed for enhancement of $T_c$ by negative $U$ charge reservoirs layers.

In the underdoped region below some not-well-defined-temperature $T^*$, well above $T_c$, anomalies are observed in various phenomena such as Knight-shifts, spin-lattice relaxation [96], transport and a reduction of the effective magnetic moments of the charge carriers. These are interpreted as crossover phenomena that we ascribe to the formation of the eh-particles that coexist with the paramagnetic doped holes. As the temperature decreases further the concentration of eh particles increases to the extent that the superconducting fluctuations observed by Ong and coworkers [97], occur, still well above $T_c$. The quasi one dimensionality of the eh-particles leads to fluctuating stripes and charge- spin separation [98]. In this model there is no necessity to postulate separate regions of (01) and (10) domains because of the d-wave symmetry that insures opposite phase relation for stripes in the (01) and (10) directions at the Cu crossing points. There would be no corresponding increase in kinetic energy because of the nodes in the $d_{x2-y2}$ wave functions at the crossing points.

However there are neutron data that indicate at least in some cases that the spin domains in the two directions are not congruent [99].

It is of interest to consider the properties of a layer in which the number of Cu sites is doubled by filling the vacant sites in the $CuO_2$ layer so as to form a cubic structure. Such a structure upon doping should have twice the superfluid density. Real space images of naturally occurring monoclinic CuO (known as tenorite) show evidence of spin-charge separation, and anisotropic transport that is consistent with stripe formation [100].

Finally, as in all models the 3d superconducting transition occurs when the temperature is lowered to $T_c$ and the 2D fluctuations condense [101]. While d-wave symmetry is favored for in the $CuO_2$ layers, a small s-wave component must exist in the chain layer cuprates a consequence of orthorhombicity, and is also likely because of disorder in all cuprates. Once a small s- component exists there is no restriction as to how large it can grow in the regions between the $CuO_2$ layers. Hence to first order there is no symmetry restriction preventing the negative-$U$ ions or ion clusters coupling with the $CuO_2$ layers and enhancing $T_c$ [102,103].

## SUMMARY

We have considered large amounts of data from the vast number of experiments concerning cuprate superconductors that have been reported over the past decade. Contrary to the commonly made assumption that interactions are confined to the $CuO_2$ planes we conclude that they are insufficient to explain the striking differences in $T_c$s that are found. We suggest that the $T_c$s found in the charge reservoir cuprates are enhanced due to superconducting pairing interactions involving the negative-$U$ ions Hg, Tl, and Bi. A striking example is the doubling of $T_c$ (from ~45K to > 90 K) found when an HgO layer is inserted into the unit cell of the 214 cuprates. Attempts to attribute this difference in $T_c$ to effects that depress the $T_c$s of the 214 cuprates are ruled out by the pressure dependence experiments that further favor our model and by other considerations as well. The collective sum of the data we have considered and interpreted makes an impressive case for the importance of negative-$U$ pairing centers.

The superconductivity found in the double chain 247 cuprates provides convincing evidence that pairing occurs outside the $CuO_2$ layers and originates in the one-dimensional chain layers. In order to account for this superconductivity and the normal state properties we hypothesize a linear diamagnetic (eh) quasi-particle that is stabilized by coulombic interactions. We speculate that this (eh) quasiparticle can exist in the $CuO_2$ layers of all the cuprates and that it offers a consistent basis for understanding the complex phase diagram of the underdoped to optimally-doped cuprates.

## Acknowledgements

We have profited from many discussions with Steven Kivelson and Boris Moyzhes, and would also like to acknowledge helpful interactions with many other colleagues including --(from A to Z) P.W. Anderson, Y. Ando, M.R. Beasley, T. Claeson, M. Greven, J. Mannhart, D. Scalapino, Y. Maeno, S. Sasaki, Z.X. Shen, H. Yamamoto, Jan Zaanen. The work has been supported by DOE and GK thanks the Netherlands Organization for Scientific Research (NWO, VENI) for support.

## Figure Captions

FIG 1. A schematic phase diagram ($T_c$ versus superfluid density $p$) that illustrates the enhancement of $T_c$ due to the insertion of charge reservoir layers that contain pairing centers. Thin dotted curve, the pairing amplitude (mean field); dotted curves; $T_{\theta 1}$ and $T_{\theta 2}$, the phase ordering temperature without and with the charge reservoir layers, respectively; Blue solid curve, $T_c$ of 214; Red dash-dot

curve, $T_c$ of 2212. As a consequence of the suppression of fluctuations the model of Kivelson and Fradkin (Fig. 4 in Kivelson's chapter in this book) would predict that $p_{opt}$ should shift to the left (lower superfluid density). The results of Karppinen et al. are taken as evidence for this shift.

FIG 2. The crystal structure of Hg-1245 (a = 3.850 Å, c = 22.126 Å) [104]. The OP undergoes the SC transition at $T_c$ = 108 K, whereas the three underdoped IP's do an AF transition below $T_N \sim 60$ K with the respective Cu moments of $\sim 0.30\mu B$ and $0.37\mu B$ at the IP and the IP. After Kotekawa et al. [43]

FIG 3. $T_{c0}$ vs $P$ for Hg-1223, Hg-1212 and Hg-1201; Inset $T_{c0}(P)$-$T_{c0}(P=0)$ for these three compounds. Open symbols taken from [105] and [106]. Arrows indicate reversibility (From L. Gao et al.. [107])

FIG 4. Pressure coefficient of $T_c$ as a function of oxygen doping for Hg-1201 (From Cao et al. [50])

FIG 5 The relationship between $T_c$ and the $CuO_2$-plane hole concentration, $p(CuO_2)$, in the $Bi_2Sr_2(Y_{1-x}Ca_x)Cu_2O_{8-d}$ system. Note that, $p(CuO_2)$ is taken as an average of the values determined for the $CuO_2$-plane hole concentration by coulometric redox analysis and by Cu $L3$-edge XANES spectroscopy. The actual cation doping level is 2 times $p(CuO_2)$. The threshold hole concentration for the appearance of superconductivity is seen at $p(CuO_2)$=0.06 (Taken from Karppinen et al. [25]).

FIG 6. Structure of $Pr_2Ba_4Cu_7O_{15-d}$ (Pr247). Pr247 consists of the Pr123 unit ("1-2-3") and the Pr124 unit ("1-2-4"). In addition to two $CuO_2$ planes, the "1-2-3" ("1-2-4") contains a single chain (a double chain). The Cu atoms in the double chain do not form a "ladder" structure but a "zigzag" chain (Taken from Sasaki *et al.*[63]).

FIG 7. (a) Temperature dependence of $1/T_1$ of the double-chain in Pr-247. Above $T_c$, the $T_1$ process exhibited a single-exponential time evolution, which yields a unique value of $T_1$. Below $T_c$, the $T_1$ process was reproduced by a bi-exponential function with two time constants, $T_1^S$ and $T_1^L$ indicating 20% of the chain copper nuclei belong to the superconducting phase. (b) Shows the antiferromagnetic magnetization of the two different copper oxide planes in the 247 unit cell (Taken from Sasaki *et al.*[63]).

FIG 8 An ionic representation of the $CuO_2$ layer in the ab plane. The squares that have no dashed lines represent the ground state of the undoped layer. Open circles—ions with filled shells either O $p^6$ and Cu $d^{10}$; filled circles –ions with open shells, O $p^5$ or Cu $d^9$. a) charge transfer exciton ; b) doped hole on oxygen c) bound exciton-hole(eh) particle; d) extended bound exciton-hole particle; e) a higher energy ($V_{pp}$) configuration of the bound exciton-hole (after Kivelson *et al.* [92]).

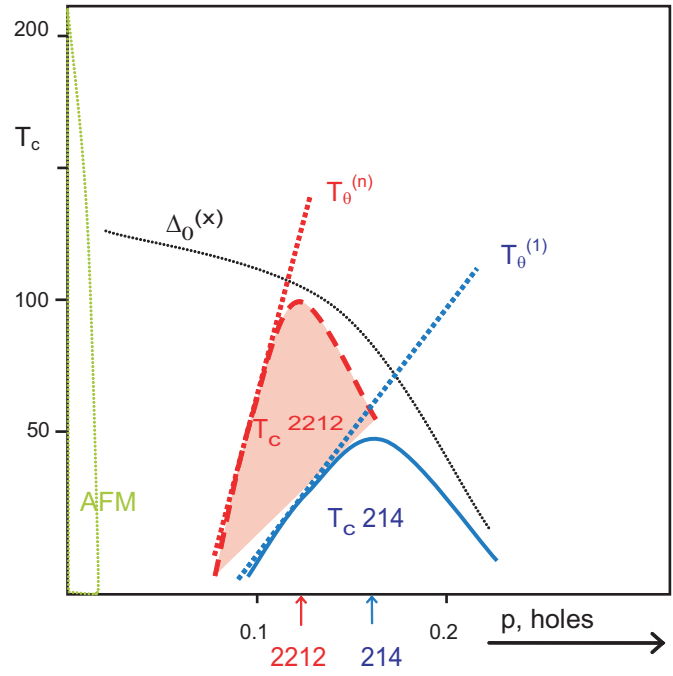

Figure 1 Geballe

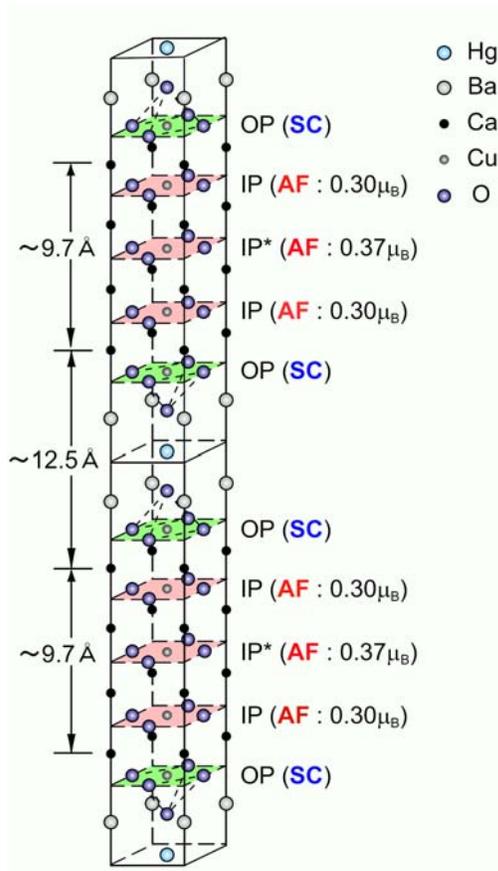

Figure 2 Geballe

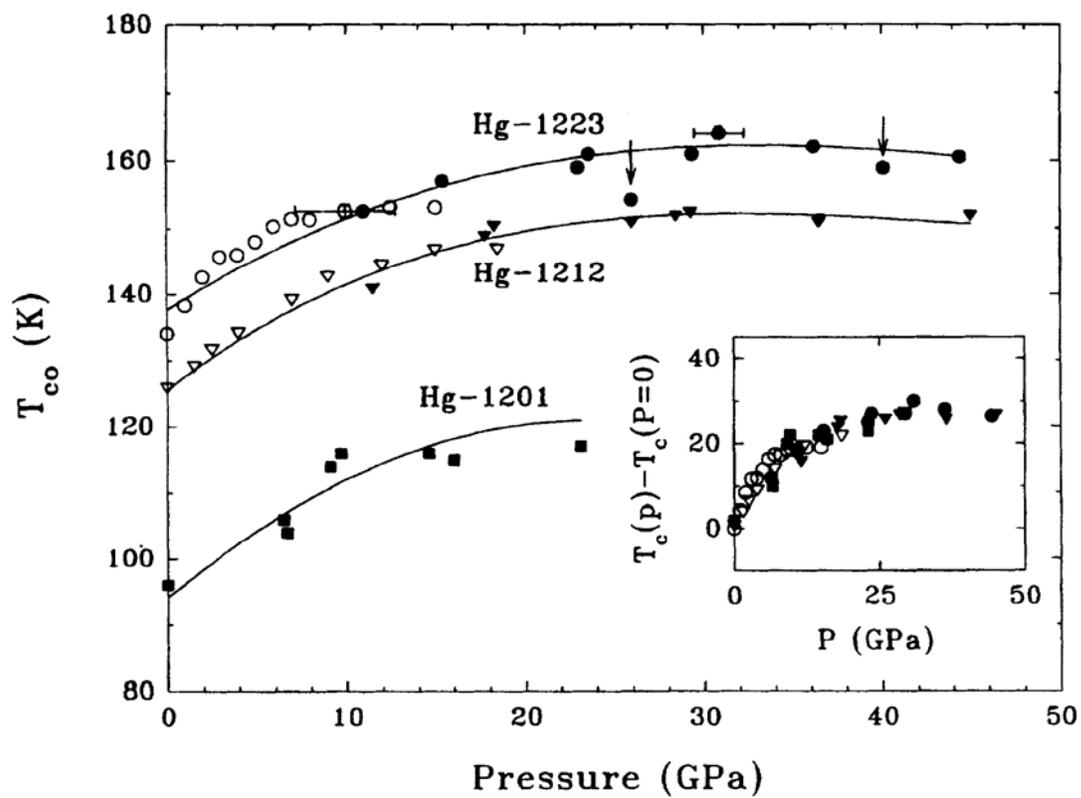

Figure 3 Geballe

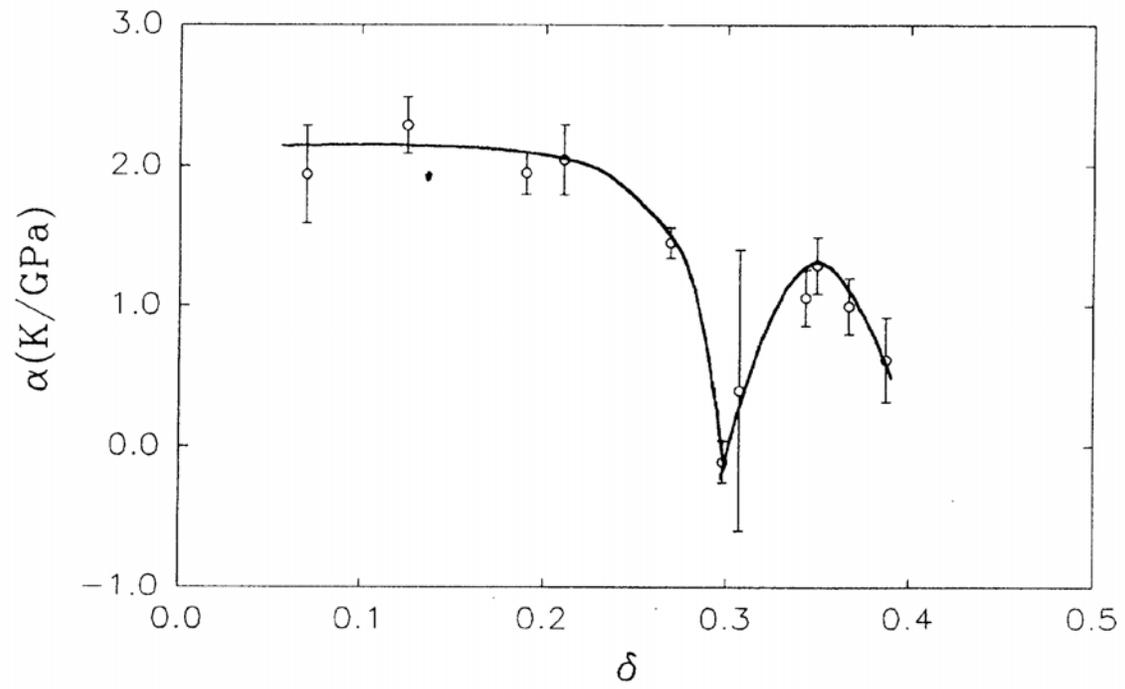

Figure 4 Geballe

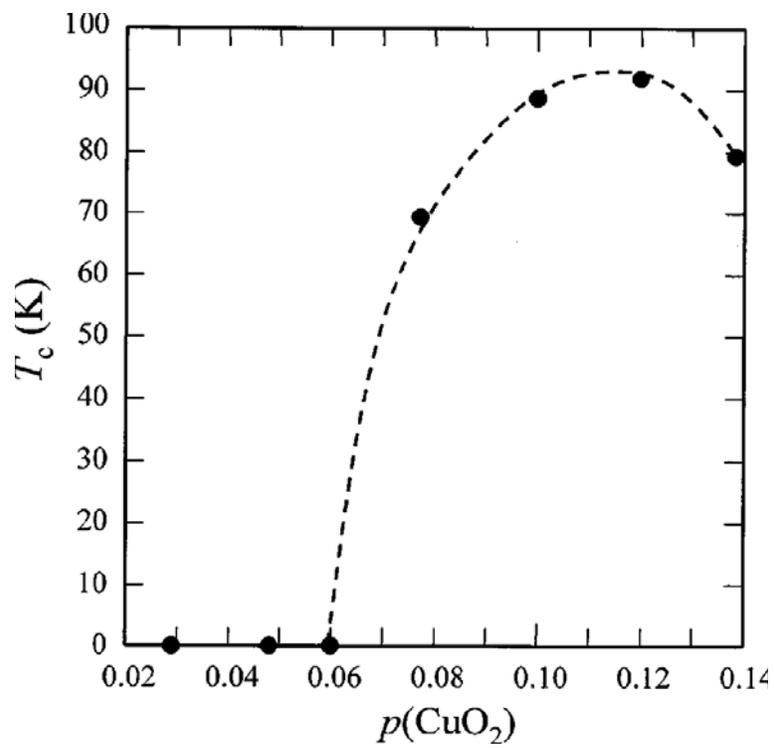

Figure 5 Geballe

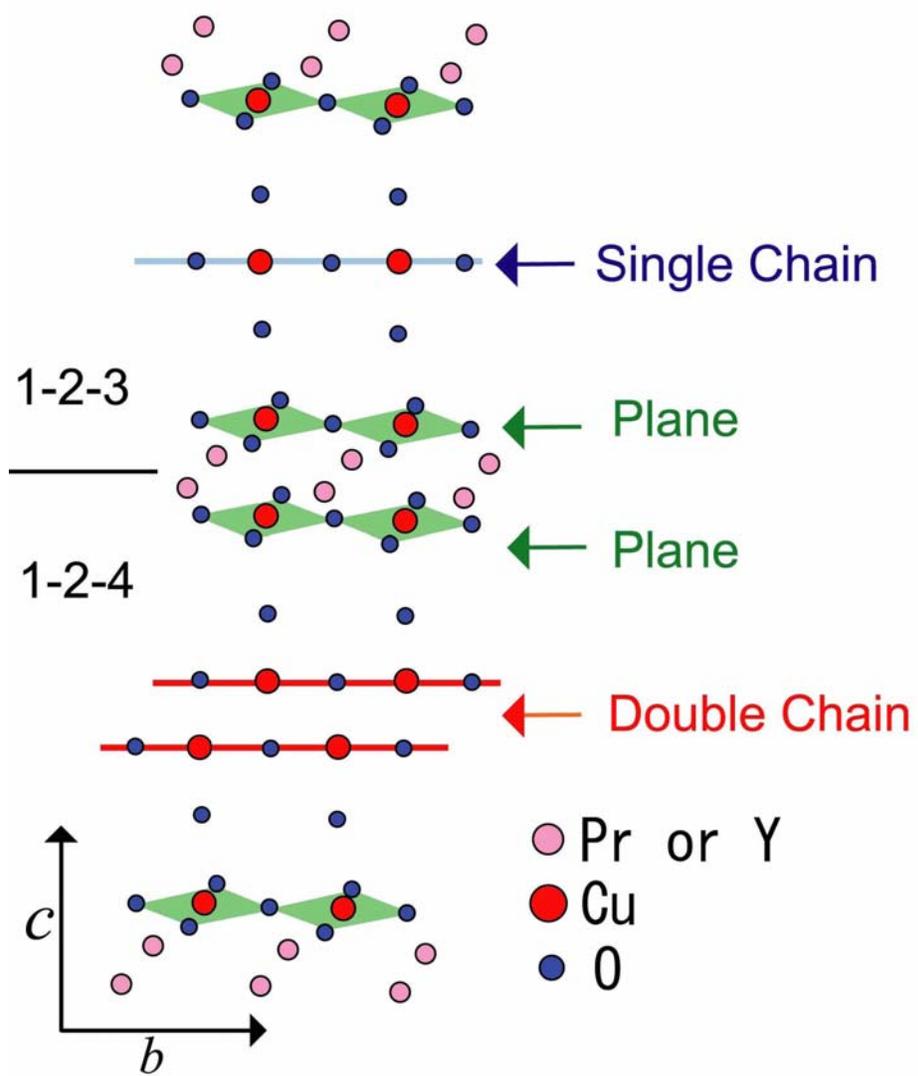

Figure 6 Geballe

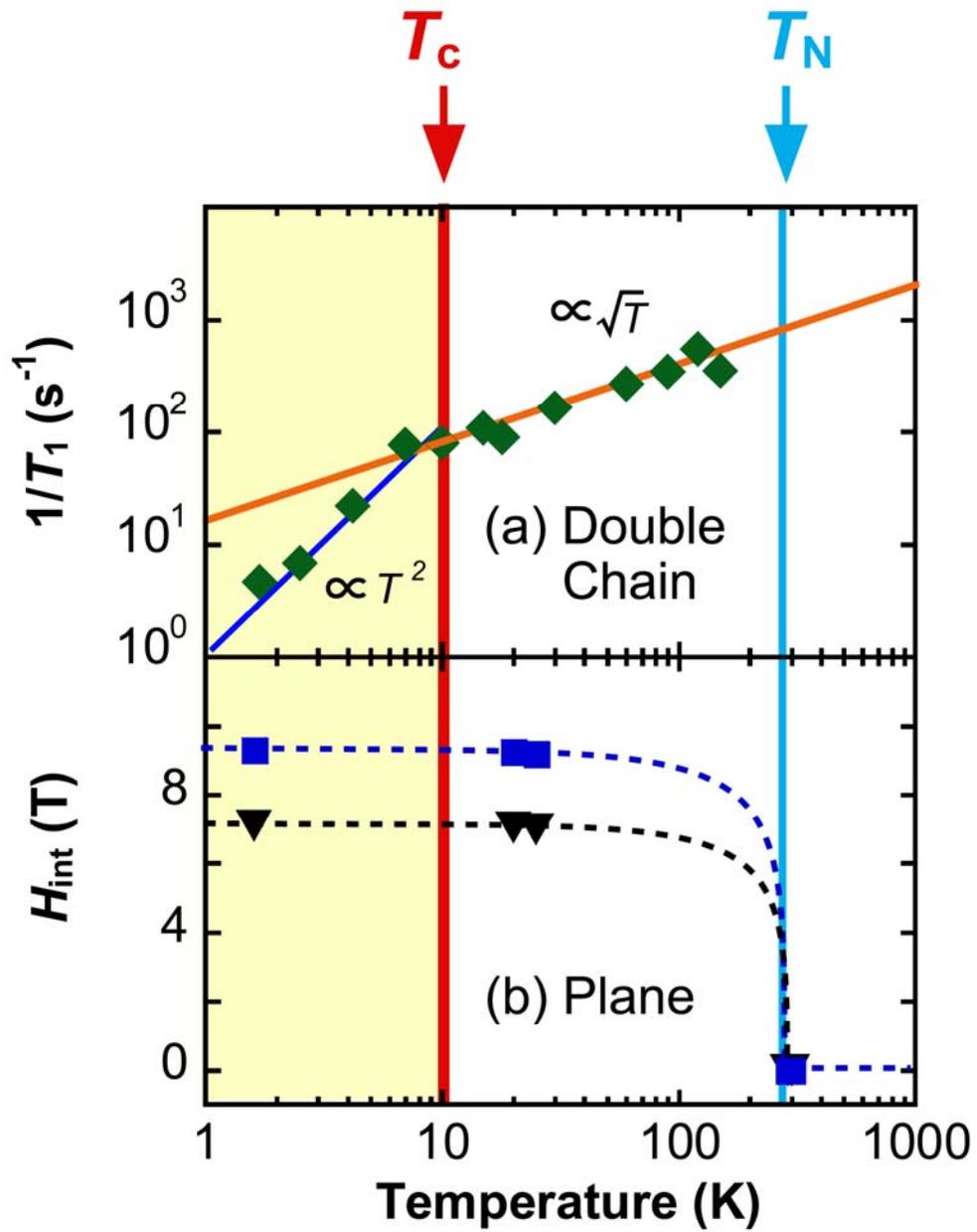

Figure 7 Geballe

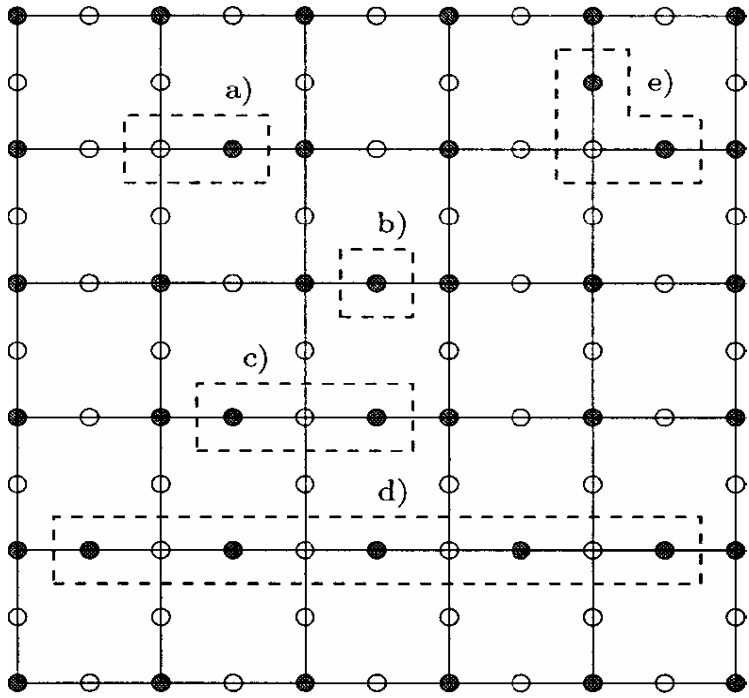

Figure 8 Geballe